\documentclass[showpacs,prd,twocolumn,amsmath,amssymb]{revtex4}
\usepackage{graphicx}
\usepackage{dcolumn}
\usepackage{bm}
\def\bra{\langle}
\def\ket{\rangle}
\def\d{\partial}
\newcommand{\R}{\mathbb{R}}
\newcommand{\cD}{\mathcal{D}}
\newcommand{\cH}{\mathcal{H}}
\newcommand{\cN}{\mathcal{N}}

\def\Ei{\mathrm{Ei}}

\def\dk#1#2{\frac{ d^{#2}{#1} }{ (2\pi)^{#2} }} 

\def\eg{{\sl e.g. }}

\def\ie{{\sl i.e. }}

\begin{document}

\title{Quantum field theory without divergences}

\author{M.V.Altaisky}
\affiliation{Joint Institute for Nuclear Research, Dubna, 141980, Russia; and Space Research Institute RAS, Profsoyuznaya 84/32, Moscow, 117997, Russia}
\altaffiliation[Also at ]{International University ``Dubna'', Universitetskaya str. 19, Dubna, 141980, Russia}
\email{altaisky@mx.iki.rssi.ru}

\date{May 5, 2013}

\begin{abstract}
It is shown that loop divergences emerging in 
the Green functions in  quantum field theory originate 
from correspondence of the Green functions to {\em unmeasurable} 
(and hence unphysical) quantities. This is because no physical 
quantity can be measured in a point, but in a region, the size 
of which is constrained by the resolution of measuring equipment. 
The incorporation of the resolution into the definition of 
quantum fields $\phi(x)\!\to\!\phi^{(A)}(x)$ and appropriate change of Feynman 
rules results in finite 
values of the Green functions. The Euclidean $\phi^4$-field theory 
is taken as an example.  
\end{abstract}

\maketitle

\section{Introduction}
The fundamental problem of quantum field theory is the problem of divergences 
of Feynman integrals. The formal infinities appearing in perturbation 
expansion of Feynman integrals are tackled with different regularization methods,
from Pauli-Villars regularization to renormalization methods for 
gauge theories, see \eg \cite{Collins1984} for a review. 
Let us consider the quantum field theory in its 
Euclidean formulation. The widely known example which fairly illustrates 
the problem is the $\phi^4$ interaction model in 
$\R^d$, see \eg \cite{Collins1984,Ramond1989}, determined by the generating functional 
\begin{equation}
W[J]= \cN \int  
e^{-\int d^dx \left[ 
\frac{1}{2}(\d\phi)^2+\frac{m^2}{2}\phi^2 + \frac{\lambda}{4!}\phi^4
 - J\phi \right]  } \cD \phi, \label{gf1}
\end{equation}
where $\cN$ is a formal normalization constant.
The connected Green functions are given 
by variational derivatives of the generating functional:
\begin{equation}
\Delta^{(n)} \equiv \bra\phi(x_1)\ldots\phi(x_n) \ket_c = 
\left. { \frac{\delta^n\ln W[J]}{\delta J(x_1) \ldots \delta J(x_n)}
}\right|_{J=0}
\label{cgf}
\end{equation}
In statistical sense these functions  have the meaning of 
the $n$-point correlation functions \cite{ZJ1999}.
The divergences of Feynman graphs in the perturbation expansion 
of the Green functions \eqref{cgf} with respect to the small 
coupling constant $\lambda$  emerge at coinciding arguments 
$x_i=x_k$. For instance, the bare two-point correlation 
function 
\begin{equation}
\Delta^{(2)}_0(x-y) = \int \frac{d^dp}{(2\pi)^d}\frac{e^{\imath p(x-y)}}{p^2+m^2}
\end{equation}
is divergent at $x\!=\!y$ for $d\ge2$.

Since in their correspondence to the $c$-valued fields the products 
$\psi^*(x)\psi(x)\Delta x$ have the probability meaning, it is quite obvious 
physically
that neither of the joint probabilities of the measured quantities can be 
infinite. The infinities seem to be caused by an  
inadequate choice of the functional space the fields belong to.

This standard approach inherited from quantum mechanics disregards 
two important notes:
\begin{enumerate}
\item To localize a particle in an interval $\Delta x$ the measuring device 
requests a momentum transfer of order $\Delta p\!\sim\!\hbar/\Delta x$. If 
the value of this momentum is too large we may get out of the applicability 
range of the initial model, in the sense that $\phi(x)$ at 
a fixed point $x$ has no experimentally verifiable meaning. What is meaningful, is the vacuum expectation of product of fields in certain 
region centered around $x$, the width of which ($\Delta x$) is constrained 
by the experimental conditions of the measurement.  
\item Even if the particle, described by the field $\phi(x)$, has been 
initially prepared on the interval 
$(x-\frac{\Delta x}{2},x+\frac{\Delta x}{2})$, the probability of 
registering this particle on this interval is generally less than unity:
for the probability of registration depends on the strength of interaction 
and the ratio of typical scales of the measured particle and the measuring 
equipment. The maximum probability of registering an object of 
typical scale $\Delta x$ by the equipment with typical resolution $a$
 is achieved when these two parameters are comparable. For this reason 
the probability of registering an electron by visual range photon scattering 
is much higher than by that of long radio-frequency waves. As 
mathematical generalization, we should say that if a measuring equipment  
with a given spatial resolution $a$ fails to register an object, prepared 
on spatial interval of width $\Delta x$ with certainty, 
then tuning the equipment to {\em all} possible resolutions $a'$ would 
lead to the registration. This certifies the fact of the existence 
of the measured object.    
\end{enumerate}

Most of the regularization methods applied to make the Green functions 
finite imply a certain type of self-similarity -- the independence of physical 
observables on the scale transformation of an arbitrary parameter of the theory --  
the cutoff length or the normalization scale. Covariance with respect 
to scale transformations is expressed by renormalization group equation 
\cite{Collins1984}. 
Another regularization idea based on self-similarity and widely used in lattice gauge theories is the Kadanoff blocking 
procedure, which averages the small-scale fluctuations up to a certain scale 
into a kind of effective interaction for a larger blocks, assuming the larger 
blocks interact with each other in the same way as their sub-blocks  
\cite{Kadanoff1966,Ito1985}. However the theory based on the Fourier 
transform of fields leaves no place for such self-similarity: the product 
of fields $\prod_i \int_{|k|<\Lambda}e^{-\imath k_i x} \tilde\phi(k_i) \dk{k}{d}$ 
describes the strength of the interaction  of all fluctuations {\em up to} the scale $1/\Lambda$, but says nothing about the interaction strength {\em at} a 
given scale. An abstract harmonic analysis based on a group $G$, which is wider than 
the group of translations $G:x\to x+b$, should be used to account for self-similarity.

The present paper aims to show how the quantum field theory of the 
scale-dependent fields can be constructed using the continuous wavelet 
transform, \ie using the decomposition of fields with respect to the representations of  
the affine group $G:x \to ax+b$. 

In {\em Section \ref{cwt:sec}} 
we present a theory of the fields $\phi_{\Delta x}(x)$, which explicitly 
depend on the resolution $\Delta x$ rather than on the point $x$ alone. 
The finiteness of the Green functions is shown on  
the simplest example of the scalar field theory with the $\phi^4$ interaction. 
In {\em Section \ref{caus:sec}} we present the commutation relations for 
the operator-valued scale-dependent fields and apply the region causality 
relations \cite{CC2005} to establish a causal ordering for scale-dependent 
fields.  Further possible applications of the proposed method, including 
that to gauge theories, and its existing discrete counterparts are mentioned 
in {\em Conclusion}.

\section{Quantum field theory based on the continuous wavelet transform \label{cwt:sec}}
To observe the two notes above we need to modify the definition of the field 
function. If the ordinary quantum field theory defines the field 
function $\phi(x)$ as a scalar product of the state vector of the system 
 and the state vector corresponding to the localization at the point $x$:
\begin{equation}
\phi(x) \equiv \bra x | \phi \ket,
\end{equation}
the modified theory should respect the resolution of the measuring 
equipment. Namely, we define the 
{\em resolution-dependent fields} 
\begin{equation}
\phi_a(x) \equiv \bra x, a; g|\phi\ket,\label{sdf}
\end{equation}
also referred to as scale components of $\phi$,
where $\bra x, a; g|$ is the bra-vector corresponding to localization 
of the measuring device around the point $x$ with the spatial resolution $a$;
$g$ labels the apparatus function of the equipment, an {\em aperture}
\cite{PhysRevLett.64.745}. In terms of the resolution-dependent 
field \eqref{sdf} the unit probability of registering the object $\phi$ 
anywhere in space at any resolution is expressed by normalization  
\begin{equation}
\int |\phi_a(x)|^2d\mu_g(a,x)=1,
\end{equation}
where $d\mu_g(a,x)$ is a translational-invariant measure, 
which depends on the position $x$, the resolution $a$, and the aperture $g$. 

Similarly to representation of a vector $|\phi\ket$ in a Hilbert space 
of states $\cH$ as a linear combination of an eigenvectors of momentum 
operator 
$
|\phi\ket=\int |p\ket dp \bra p |\phi\ket,$
any $|\phi\ket \in \cH$ can be represented as a linear combination 
of different scale components 
\begin{equation}
|\phi\ket=\int_G |g;a,b\ket d\mu(a,b)\bra g;a,b|\phi\ket. \label{gwl} 
\end{equation}
Here, according to \cite{Carey1976,DM1976},  $|g;a,b\ket=U(a,b)|g\ket$; 
$d\mu(a,b)$ is the left-invariant measure on the affine group $G$;
$U(a,b)$ is a representation of the affine group $G:x'=ax+b$; 
$|g\ket\in\cH$ is 
a {\em admissible vector}, satisfying the condition
$$
C_g = \frac{1}{\| g \|^{2}} \int_G |\bra g| U(a,b)|g \ket |^2 
d\mu(a,b)
<\infty. 
$$

If the measuring equipment has the resolution $A$, \ie all 
states $\bra g;a\ge A,x|\phi\ket$ are registered with significant 
probability, but those with $a<A$ are not, the regularization of the 
model \eqref{gf1} in momentum space, 
with the cutoff momentum $\Lambda=2\pi/A$ corresponds to the 
UV-regularized functions
\begin{equation}
\phi^{(A)}(x) =\frac{1}{C_g}\int_{a\ge A} 
\bra x|g;a,b\ket d\mu(a,b)\bra g;a,b|\phi\ket.
\end{equation}
The regularized $n$-point Green functions are 
$
\mathcal{G}^{(A)}(x_1,\ldots,x_n) \equiv \bra\phi^{(A)}(x_1),\ldots, \phi^{(A)}(x_n) \ket_c 
.$

However, the momentum cutoff is merely a technical trick: the physical 
analysis, performed by renormalization group method \cite{Wilson1973,Collins1984}, demands 
the independence of physical results from the cutoff at $\Lambda\to\infty$.

In present paper we give an alternative, geometrical, interpretation to 
the cutoff. We assert that {\sl if for a given physical system $\phi$ and 
given measuring equipment there 
exist the finest resolution scale $A$, so that it is impossible to measure 
any physical quantity related to $\phi$ with a resolution $a<A$, then any 
description of $\phi$ should comprise only such functions, the typical 
variation scales of which are not less than $A$}. This looks like we observe 
the system $\phi$ {\em from outside the scale $A$}. The Feynman functional 
integrations in this approach are performed only over the functions with 
typical scales $a\!\ge\!A$. Our method does not apply any direct cutoff to 
the momenta -- the arguments of the Fourier transform. The momentum 
conservation in each vertex remains intact. The 
calculations can be performed either for the scale-component Green functions 
$\bra\phi_{a_1}(x_1)\ldots \phi_{a_n}(x_n) \ket$, or for the integrals 
of those over the scales $\bra\phi^{(A_1)}(x_1)\ldots \phi^{(A_n)}(x_n) \ket$.

The technical realization of our scheme is based on the substitution 
of the fields $\bra x|\phi\ket$ with $|\phi\ket$ given by \eqref{gwl} 
into the generating functional \eqref{gf1}. In coordinate representation 
this is known as the continuous wavelet transform (see \eg 
\cite{Daub10}).
To keep the scale-dependent fields $\phi_a(x)$
the same physical dimension as the ordinary fields $\phi(x)$ we write the coordinate
representation of wavelet transform in $L^1$-norm \cite{PhysRevLett.64.745,Chui1992,HM1998}:
\begin{eqnarray}
\phi(x) &=& \frac{1}{C_g} \int \frac{1}{a^d} g\left(\frac{x-b}{a}\right) \phi_a(b) \frac{dad^db}{a},
\label{iwt} \\
\phi_a(b) &=& \int \frac{1}{a^d} \overline{g\left(\frac{x-b}{a}\right)} \phi(x) d^dx.
\label{dwt}
\end{eqnarray}
In the latter equations the field $\phi_a(b)$ has a physical meaning of 
the amplitude of the field $\phi$ measured at point 
$b$ using a device with an aperture $g$ and a tunable spatial resolution $a$.
For isotropic wavelets $g$ the normalization 
constant
$C_\psi$ is readily evaluated using Fourier transform:
\begin{equation}
C_g = \int_0^\infty |\tilde g(ak)|^2\frac{da}{a}
= \int |\tilde g(k)|^2 \frac{d^dk}{S_{d}|k|}<\infty,
\label{adcf}
\end{equation}
where $S_d = \frac{2 \pi^{d/2}}{\Gamma(d/2)}$ is the area of unit sphere 
in $\R^d$.

Substitution of the continuous wavelet transform \eqref{iwt} into field theory \eqref{gf1}
gives the generating functional for the scale-dependent fields $\phi_a(x)$ 
\cite{AltSIGMA07}:
\begin{widetext}
\begin{eqnarray} \nonumber 
W_W[J_a] &=&\cN \int \cD\phi_a(x) \exp \Bigl[ -\frac{1}{2}\int \phi_{a_1}(x_1) D(a_1,a_2,x_1-x_2) \phi_{a_2}(x_2)
\frac{da_1d^dx_1}{a_1}\frac{da_2d^dx_2}{a_2}  \\
&-&\frac{\lambda}{4!}
\int V_{x_1,\ldots,x_4}^{a_1,\ldots,a_4} \phi_{a_1}(x_1)\cdots\phi_{a_4}(x_4)
\frac{da_1 d^dx_1}{a_1} \frac{da_2 d^dx_2}{a_2} \frac{da_3 d^dx_3}{a_3} \frac{da_4 d^dx_4}{a_4} 
+ \int J_a(x)\phi_a(x)\frac{dad^dx}{a}\Bigr], \label{gfw}
\end{eqnarray}
\end{widetext}
with $D(a_1,a_2,x_1-x_2)$ and $V_{x_1,\ldots,x_4}^{a_1,\ldots,a_4}$ denoting the wavelet images of the inverse propagator and that of the interaction potential. 
The Green functions for scale component fields are given by 
functional derivatives
$$
\bra\phi_{a_1}(x_1)\cdots\phi_{a_n}(x_n)\ket_c
= \left. \frac{\delta^n \ln W_W[J_a]}{\delta J_{a_1}(x_1)\ldots 
\delta J_{a_n}(x_n)} \right|_{J=0}.$$
Surely the integration in \eqref{gfw} over all 
scale variables $\int_0^\infty \frac{da_i}{a_i}$ turns us back 
to the divergent theory \eqref{gf1}.

This is the point to restrict the functional integration in \eqref{gfw} 
only to the field configurations $\{ \phi_a(x) \}_{a\ge A}$. The restriction 
is imposed at the level of the Feynman diagram technique. Indeed, applying 
the Fourier transform to the r.h.s. of (\ref{iwt},\ref{dwt}) one yields 
\begin{eqnarray*}
\phi(x) &=& \frac{1}{C_g} \int_0^\infty \frac{da}{a} \int \dk{k}{d} e^{-\imath k x}
\tilde g(ak) \tilde \phi_a(k), 
\\  
\tilde\phi_a(k) &=& \overline{\tilde g(ak)}\tilde\phi(k) .
\end{eqnarray*}
Doing so, we have the following modification of the Feynman diagram technique
\cite{Alt2002G24J}:
\begin{itemize}\itemsep=0pt
\item each field $\tilde\phi(k)$ will be substituted by the scale component
$\tilde\phi(k)\to\tilde\phi_a(k) = \overline{\tilde g(ak)}\tilde\phi(k)$.
\item each integration in momentum variable is accompanied by corresponding 
scale integration:
\[
 \dk{k}{d} \to  \dk{k}{d} \frac{da}{a}.
 \]
\item each interaction vertex is substituted by its wavelet transform; 
for the $N$-th power interaction vertex this gives multiplication 
by factor 
$\displaystyle{\prod_{i=1}^N \overline{\tilde g(a_ik_i)}}$.
\end{itemize}
The finiteness of the loop integrals is provided by the following rule:
{\bf there should be no scales $a_i$ in internal lines smaller than the minimal scale 
of all external lines}. Therefore the integration in $a_i$ variables is performed from 
the minimal scale of all external lines up to the infinity.

To illustrate the method we present the calculation of the one-loop contribution to the 
two- and the four-point Green functions in $\phi^4$ model in $\R^4$. The best choice of 
the wavelet function $g(x)$ would be the apparatus function of 
the measuring device, however a simple choice   
\begin{equation}
g(x)=-x e^{-x^2/2},\quad \tilde g(k) = (-\imath k) e^{-k^2/2}
\label{g1}
\end{equation}
demonstrates the method qualitatively. The function \eqref{g1} is well localized in 
both the coordinate and the momentum spaces, it satisfies the admissibility condition with $C_g=1$. Due to the property $\int_{-\infty}^\infty g(x)dx=0$ the detector 
with such aperture is insensitive to constant fields, but detects the 
gradients of the fields. 

\begin{figure}[ht]
\includegraphics[width=.33\textwidth]{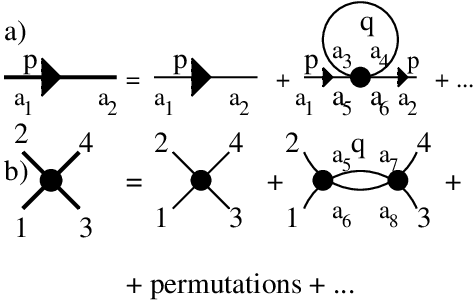}
\caption{Feynman diagrams for the Green functions $G^{(2)}$ and $G^{(4)}$ 
for the resolution-dependent fields}
\label{gf:pic}
\end{figure}

Let us consider the contribution of the tadpole diagram to the two-point 
Green function $G^{(2)}(a_1,a_2,p)$ shown in 
Fig.~\ref{gf:pic}a. The bare Green function is 
\begin{equation}
G^{(2)}_0(a_1,a_2,p) = \frac{\tilde g(a_1p)\tilde g(-a_2p)}{p^2+m^2}.
\end{equation}   
The tadpole integral, to keep with the notation of \cite{AltSIGMA07}, 
is written as  
\begin{eqnarray*}
T_1^d(Am)&=&\frac{1}{C_g^2} \int_{a_3,a_4 \ge A}  \frac{d^dq}{(2\pi)^d} 
\frac{|\tilde g(a_3 q)|^2 |\tilde g(-a_4 q)|^2}{q^2+m^2}
 \frac{da_3}{a_3} \frac{da_4}{a_4}  
\\
&=& 
\frac{S_d m^{d-2}}{(2\pi)^d}\int_0^\infty f^2(Amx)\frac{x^{d-1}dx}{x^2+1} \\
f(x) &\equiv& \frac{1}{C_g}\int_x^\infty |\tilde g(a)|^2\frac{da}{a}.
\end{eqnarray*}
For our simple model aperture \eqref{g1} the filtering factor is  
$f(x)=e^{-x^2}.$

In $d=4$ dimension we get 
\begin{equation}
T^4_1(\alpha) = \frac{-4\alpha^4 e^{2\alpha^2} \Ei(1,2\alpha^2)+2\alpha^2
}{
64\pi^2\alpha^4}m^2, 
\label{t41}
\end{equation}
where $\alpha\!\equiv\!Am$ is dimensionless scale factor, 
$A\!=\!\min(a_1,a_2)$, and 
$$\Ei(1,z)=\int_1^\infty \frac{e^{-xz}}{x}dx$$ denotes the exponential integral.
Finally, the $O(\lambda)$ contribution to the two-point Green function in 
$\R^d$, shown in Fig.~\ref{gf:pic}a, is 
\begin{widetext}
\begin{eqnarray}
G^{(2)}(a_1,a_2,p) &=& \frac{\tilde g(a_1p)\tilde g(-a_2p)}{p^2+m^2} - 
\label{g21} 
\frac{\lambda}{2}
\frac{\tilde g(a_1p)\tilde g(-a_2p) f^2(Ap)T^d_1(Am) }{\left(p^2+m^2\right)^2}
+ \ldots .
\end{eqnarray}
\end{widetext}
In the one-loop contribution to the vertex, shown in Fig.~\ref{gf:pic}b,
the value of the loop integral is 
\begin{equation}
X_d = \frac{\lambda^2}{2}\frac{1}{(2\pi)^d}\int \frac{d^dq}{(2\pi)^d}
\frac{f^2(qA)f^2((q-s)A)}{\left[ q^2+m^2\right]\left[ (q-s)^2+m^2\right] },
\label{I4}
\end{equation}
where $s\!=\!p_1\!+\!p_2, A=\min(a_1,a_2,a_3,a_4)$. The integral \eqref{I4} can 
be calculated by symmetrization of loop momenta $q\!\to\!q\!+\!\frac{s}{2}$ in 
Fig.~\ref{gf:pic}b, 
doing so  after a simple algebra  we yield   
\begin{eqnarray}\nonumber
X_d&=&\frac{\lambda^2}{2}\frac{S_{d-1}}{(2\pi)^{2d}}s^{d-4}e^{-A^2s^2} 
\int_0^\infty e^{-4A^2s^2y^2}I_d(y)y^{d-3}dy, \\
 I_d(y)&=&\int_0^\pi \frac{\sin^{d-2}\theta d\theta}{\beta^2(y)-\cos^2\theta}, 
\quad \beta(y) = \frac{y^2+\frac{1}{4}+\frac{m^2}{s^2}}{y}, \label{xd}
\end{eqnarray}
where $\theta$ is the angle between the loop momentum $q$ and the total 
momentum $s$. 

\begin{figure}[ht]
\includegraphics[width=.33\textwidth]{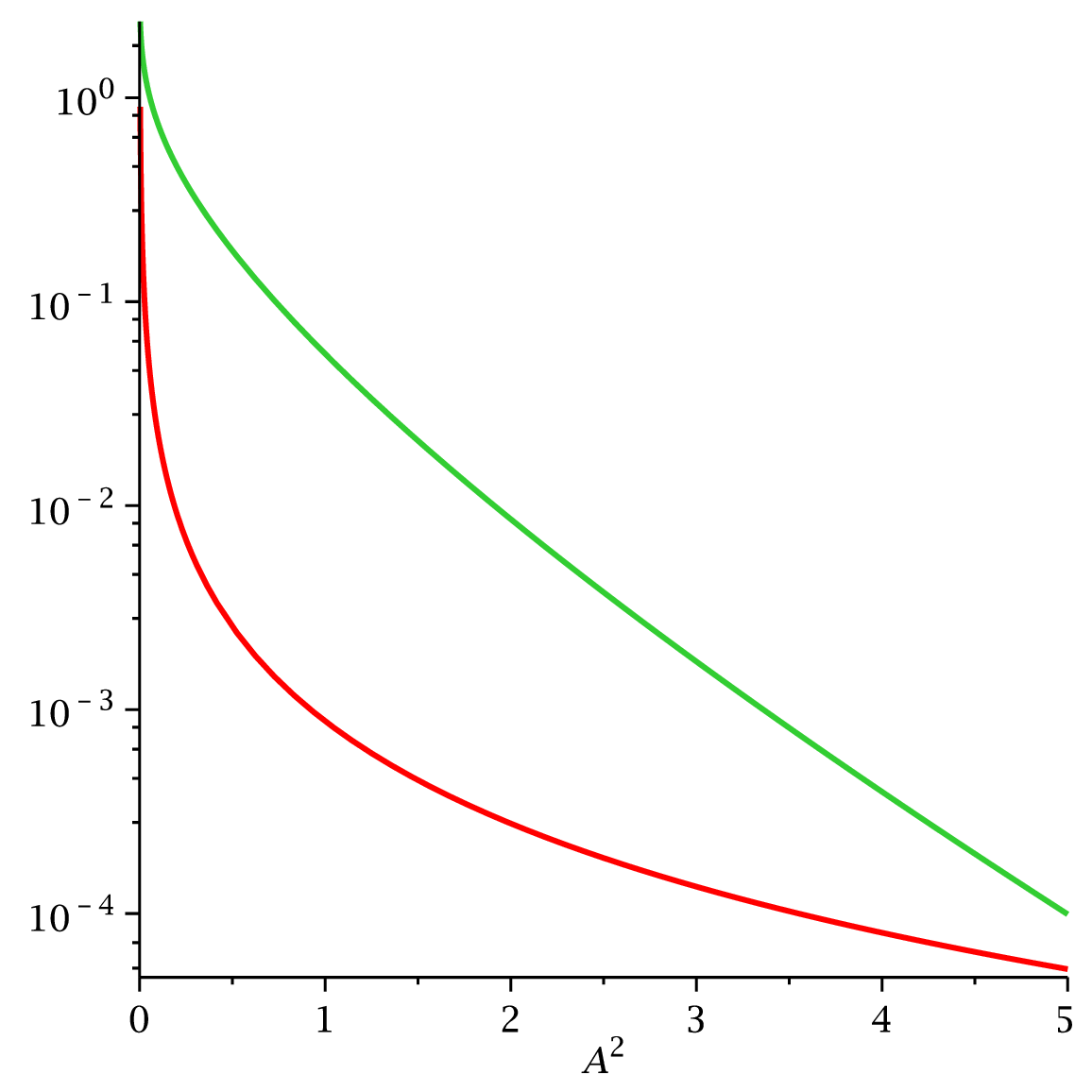}
\caption{Scale-decay factors for the two-point and four-point Green 
functions. The bottom curve is 
the graph of \eqref{t41} as a function of $A^2$; the top curve 
is the graph of \eqref{f4ass} divided by $\frac{\lambda^2}{256\pi^6}$ 
as a function of $A^2$. $m=s^2=1$ is set for both curves
}
\label{cut:pic}
\end{figure}

In critical dimension $d=4$
\begin{equation}
X_4 = \frac{\lambda^2 }{256\pi^6}e^{-A^2s^2}\int_0^\infty 
e^{-4A^2s^2y^2}\left(1-\sqrt{1-\beta^{-2}(y)} \right)dy^2.
\label{x4d}
\end{equation}
In Fig.~\ref{cut:pic} below we present the graph of large momentum 
asymptotics of \eqref{x4d}
\begin{widetext} 
\begin{eqnarray}
\lim_{s^2\gg 4m^2} X_4(\alpha^2) = \frac{\lambda^2}{256\pi^6} \frac{e^{-2\alpha^2}}{2\alpha^2} 
\bigl[e^{\alpha^2}-1 
-\alpha^2e^{2\alpha^2}\Ei(1,\alpha^2)
+ 2\alpha^2e^{2\alpha^2}\Ei(1,2\alpha^2)
  \bigr], \label{f4ass}
\end{eqnarray}
\end{widetext}
where $\alpha\equiv As$, 
compared to the \eqref{t41} factor of the two-point Green function.
Other diagrams contributing to the vertex shown in Fig.~\ref{gf:pic}b give similar  
factors with appropriate substitution of $s$ to $s=p_i+p_j$.

Turning back to the coordinate representation of the Green functions 
for the fields $\phi_a(x)$, we can see there no 
divergences 
at coinciding spatial arguments. Say, the bare two-point Green function
$$ G^{(2)}_0(a_1,a_2,b_1-b_2)=\int \frac{d^dp}{(2\pi)^d}e^{\imath p(b_1-b_2)}
\frac{\tilde g(a_1p)\tilde g(-a_2p) }{p^2+m^2}$$ gives at our 
model choice \eqref{g1} in $d\!=\!4$ dimension  
\begin{widetext}
\begin{eqnarray*}
G^{(2)}_0(a_1,a_2,b_1-b_2=0)= \pi^2 m^2 \alpha_1\alpha_2 
\left[\frac{4}{\left(\alpha_1^2+\alpha_2^2\right)^2} 
-\frac{2}{\alpha_1^2+\alpha_2^2}+e^\frac{\alpha_1^2+\alpha_2^2}{2}
\Ei \left(1,\frac{\alpha_1^2+\alpha_2^2}{2}\right)
 \right],
\end{eqnarray*}
\end{widetext}
where $\alpha_i\!=\!a_i m$ are dimensionless scale parameters. 

We would like to emphasize that in spite of the fact 
that application of wavelets to quantum field theory is not new, 
the  interpretation 
of the fields $\phi_a(x)$ (or their integrals $\phi^{(A)}$) as physical fields
yields a finite theory with no need for renormalization. 
Indeed, the t'Hooft and Veltman dimensional regularization scheme 
\cite{HooftVeltman1972} works perfectly well in the presence of 
the scale factor $A$. The difference is that the integrated function $f$ 
in 
\begin{equation}
I_A = \int d^4\underline{p}\int_0^\infty 
d\omega \omega^{n-5}\frac{2\pi^{\frac{n}{2}-1}}{\Gamma\left(\frac{n}{2}-2 \right)}
f(A,\underline{p},\omega^2)\label{hv},
\end{equation}
where $n$ is the formal integration dimension, in our case contains 
the exponential factor $f(A,\underline{p},\omega^2) \sim
\exp(-A^2({\underline{p}}^2+\omega^2))$, which suppress all ultraviolet divergences.
In the limit $A\to0$ the integration by parts in \eqref{hv} over the $\omega^2$ 
argument recovers the well known poles at the physical dimension $n=4$.

\section{Causality and commutation relations \label{caus:sec}}
We have considered a multiscale scalar field theory 
determined by the generating functional \eqref{gfw}. Such theory is 
used if the field $\phi_a(x)$ is a {\em $c$-valued} function. In quantum 
field theory adjusted to high energy physics applications, the fields 
$\phi_a(x)$ are operator-valued functions. So, as it was 
already emphasized in the context of the wavelet application to 
quantum chromodynamics \cite{Federbush1995}, the operator ordering and the commutation 
relations  are to be defined.

The commutation relations $[\phi_a(x),\phi_{a'}(x')]$    can be 
imposed in such a way that they recover ordinary commutation relations after 
integration 
over the scale arguments. This was already done in \cite{AltaiskyPEPAN2005}. 
The decomposition of the operator-valued field $\hat \phi(x)$ into 
the positive and negative frequency scale components is 
\begin{equation}
\hat \phi(x)=\int\frac{da}{a}\dk{k}{d} \frac{\tilde g(ak)}{C_g} \left[
e^{\imath k x} u^+_a(k)+e^{-\imath k x}u^-_a(k)
\right],
\end{equation}
where $u^\pm_a(k)=u_a(\pm k)\theta(k_0)$. 
Since 
$$
u^\pm(k) = \frac{1}{C_g}\int \frac{da}{a}\tilde g(ak)u^\pm_a(k),
$$
the standard commutation relations can be satisfied if we set 
\begin{equation}
[u^+_{a_1}(k_1),u^-_{a_2}(k_2)]=C_g a_1 \delta(a_1-a_2)[u^+(k_1),u^-(k_2)].
\end{equation}

As was shown in \cite{AE1974}, the non-local field 
theory with the propagator cutoff $V(l^2k^2)$ satisfies the 
{\em microcausality condition} for the $S$ matrix \cite{Bog1955} 
\begin{equation}
\frac{\delta}{\delta \phi(x)} \left(\frac{\delta S}{\delta \phi(y)}S^+ \right)=0
\quad \hbox{for\ } x \stackrel{<}{\sim} y \label{mcaus}
\end{equation}
in each order of the perturbation theory. For the theory of scale-dependent 
fields, a stronger microcausality condition 
\begin{equation}
\frac{\delta}{\delta \phi_a(x)} \left(\frac{\delta S}{\delta \phi_b(y)}S^+ 
\right)=0
\quad \hbox{for\ } x \stackrel{T}{<} y \quad \hbox{or\ } x \sim y \label{mcaus1}
\end{equation} 
may be suggested if the derivation is performed 
with the generalized causal $T$-ordering ("the coarse acts first'') 
defined in \cite{AltaiskyPEPAN2005} according to the region causality rules
\cite{CC2005}.
The definition of the generalized causal ordering given in 
\cite{AltaiskyPEPAN2005}, is the following:  
\begin{equation}
T( A_{\Delta x}(x) B_{\Delta y}(y) ) = \begin{cases}
A_{\Delta x}(x) B_{\Delta y}(y), & y_0 < x_0, \\
\pm B_{\Delta y}(y) A_{\Delta x}(x), & x_0 < y_0, \\
A_{\Delta x}(x) B_{\Delta y}(y), & \Delta x \subset \Delta y, \\
\pm B_{\Delta y}(y) A_{\Delta x}(x), & \Delta y \subset \Delta x,
\end{cases}
\label{tnew}
\end{equation}
\ie, if the region $\Delta x$ is inside the region $\Delta y$, the operator 
related to the larger region $\Delta y$ acts on vacuum first. If the 
regions $\Delta x$ and $\Delta y$ (the vicinities of two distinct points 
$x\ne y$) have zero intersection $\Delta x \cap \Delta y=\emptyset$, the causal 
ordering \eqref{tnew} coincides with usual $T$-ordering.

\section{Conclusion}
In this paper we presented a regularization method for quantum field 
theory based on the continuous wavelet transform. The idea of substituting 
wavelet decomposition of the fields into the action functional is not 
new. It was used by many authors, but using the {\em discrete} wavelet 
transform. This efficiently works for the Monte Carlo simulations \cite{HS1995,Best2000}, and provides a frame for renormalization 
\cite{BF1983,BF1987}, including the regularization of gauge theories 
\cite{Federbush1995}. In many aspects, the discrete wavelet transform 
works as a lattice regularization \cite{Battle1989}.
The novelty of the approach presented in this paper consists in 
using the {\em continuous} wavelet transform of the fields (along with the 
region causality assumptions \cite{CC2005}) with the operator 
ordering rules given in \cite{AltaiskyPEPAN2005}.

An attempt to apply the continuous wavelet transform 
to the $\phi^4$ field theory was undertaken in \cite{ffp4} 
based on the general ideas of the wavelet transform on the Poincare group
\cite{KlaStre91}. However, a physical interpretation of the wavelet 
transform scale argument as a physical parameter of observation was 
given much later in \cite{AltaiskyPEPAN2005,AltSIGMA07} in the context 
of quantum electrodynamics. The key issue of the quantum field theory  
is gauge invariance. In our wavelet framework, this problem was 
addressed in \cite{AA2009}, where the Ward-Takahashi identities for 
$U(1)$ gauge theory were derived. Later, we are going to consider this problem in more detail.

\begin{acknowledgments}
The author is thankful to prof. N.V.Antonov for critical 
comments, and to profs. A.E.Dorokhov and V.B.Priezzhev for useful discussions.
The research was supported in part by DFG Project 436 RUS 113/951.
\end{acknowledgments}


\end{document}